\documentclass{ws-procs975x65}

\begin{document}

\title{Main parameters of neutron stars from quasi-periodic oscillations \\ in low mass X-ray binaries}

\author{Kuantay Boshkayev,$^{1,2,*}$  Jorge A. Rueda,$^{2}$ and  Marco Muccino.$^{2}$}

\address{$^1$Faculty of Physics and Technology, IETP, Al-Farabi Kazakh National University,\\
Al-Farabi avenue 71, Almaty, 050040, Kazakhstan\\
$^2$International Center for Relativistic Astrophysics Network,\\
Piazza della Repubblica 10, Pescara, I-65122, Italy\\
$^*$E-mail: kuantay@mail.ru}

\begin{abstract}
We investigate the kilohertz quasi-periodic oscillations of low-mass X-ray binaries within the Hartle-Thorne spacetime. On the basis the relativistic precession model we extract the total mass $M$, angular momentum $J$, and quadrupole moment $Q$ of a compact object in a low-mass X-ray binary by analyzing the data of the Z -source GX 5-1. In view of the recent neutron star model we compute the radius, angular velocity and other parameters of this source by imposing the observational and theoretical constraints on the mass-radius relation. 
\end{abstract}



\bodymatter

\

\section{Introduction}
In our recent work using the Hartle-Thorne external solution we have derived the azimuthal (Keplerian), radial and polar (vertical) fundamental frequencies.\cite{boshkayevqpos2014} The application of these frequencies to the observed quasiperiodic oscillations (QPO) from the low-mass X-ray binaries has been considered on the basis of the relativistic precession model (RPM).\cite{ste-vie:1998,ste-vie:1999,ste-etal:1999} We examined rotating spacetimes that comprehended the effects of frame-dragging and quadrupolar deformation of the source and fit directly the relation between the twin-peak quasiperiodic frequencies.\cite{boshkayevqpos2015} We showed that a statistically preferred fit is obtained for the case of three parameters: mass $M$, angular momentum $J$ and quadrupole moment $Q$ with respect to an analysis using only $M$ and $J$.\cite{2010ApJ...714..748T}


In this work, we extend  our previous results considering the equilibrium structure of rotating neutron stars within the model proposed in Ref. \refcite{belvedere2012} by means of both interior and exterior Hartle-Thorne solutions.\cite{belvedere2014} By fulfilling all the stability criteria and the latest observational and theoretical constraints on neutron star mass-radius relations, we compute the mass, radius, rotation frequency and other parameters of the Z -source GX 5-1.


Our paper is organized as follows: in Section \ref{sec:2}, we consider the external Hartle-Thorne solution; in Section \ref{sec:3}, we discuss about the extraction of the $M$, $J$ and $Q$ of GX 5-1 from its QPO data; in Section  \ref{sec:4}, we consider observational and theoretical constraints on the mass-radius relations of neutron stars and in Section \ref{sec:5} we calculate the main physical parameters of GX 5-1 such as radius and rotational frequency. Finally, in Section \ref{sec:6}, we summarize our main results, discuss their significance, and draw our conclusions.

%
\section{The Hartle-Thorne metric}\label{sec:2}

The Hartle-Thorne metric \cite{har-tho:1968,boshkayev2012} describing the
exterior field of a slowly rotating slightly deformed object is given by
\begin{eqnarray}\label{ht1}
ds^2&=&\left(1-\frac{2{ M }}{r}\right)\left[1+2k_1P_2(\cos\theta)+2\left(1-\frac{2{ M}}{r}\right)^{-1} \frac{J^{2}}{r^{4}}(2\cos^2\theta-1)\right]dt^2
\nonumber\\ && -\left(1-\frac{2{ M}}{r}\right)^{-1}\left[1-2\left(k_1-\frac{6 J^{2}}{r^4}\right)P_2(\cos\theta) -2\left(1-\frac{2{ M}}{r}\right)^{-1}\frac{J^{2}}{r^4}\right]dr^2
\nonumber\\ &&-r^2[1-2k_2P_2(\cos\theta)](d\theta^2+\sin^2\theta d\phi^2)+\frac{4J}{r}\sin^2\theta dt d\phi,
\end{eqnarray}
where
\begin{eqnarray}\label{ht2}
k_1&=&\frac{J^{2}}{{ M}r^3}\left(1+\frac{{ M}}{r}\right)+\frac{5}{8}\frac{Q-J^{2}/{ M}}{{ M}^3}Q_2^2\left(x\right)\ ,\nonumber \\
k_2&=&k_1+\frac{J^{2}}{r^4}+\frac{5}{4}\frac{Q-J^{2}/{ M}}{{ M}^2r}\left(1-\frac{2{ M}}{r}\right)^{-1/2}Q_2^1\left(x\right)\ ,\nonumber
\end{eqnarray}
and
\begin{eqnarray}\label{legfunc2}
Q_{2}^{1}(x)&=&(x^{2}-1)^{1/2}\left[\frac{3x}{2}\ln\frac{x+1}{x-1}-\frac{3x^{2}-2}{x^{2}-1}\right],\nonumber \\
Q_{2}^{2}(x)&=&(x^{2}-1)\left[\frac{3}{2}\ln\frac{x+1}{x-1}-\frac{3x^{3}-5x}{(x^{2}-1)^2}\right],
\end{eqnarray}
are the associated Legendre functions of the second kind, with $x=r/M -1$, and $P_2(\cos\theta)=(1/2)(3\cos^2\theta-1)$ is the Legendre polynomial. The constants ${M}$, ${J}$ and ${Q}$ are the total mass, angular momentum and  quadrupole moment of a rotating object, respectively.

The Hartle-Thorne metric is an approximate solution of vacuum Einstein field equations that describes the exterior of any slowly and rigidly rotating, stationary and axially symmetric body. The metric is given with accuracy up to the second order terms in the body's angular momentum, and first order in its quadrupole moment. Unlike other solutions of the Einstein field equations this solution possesses its internal counterpart and valid in the strong field regime with intermediate rotation rate,\cite{ber-etal:2005} which is quite enough for our analyses.
%
%
\section{Extraction of the Mass, Angular Momentum and Quadrupole Moment}\label{sec:3}
Spacetimes around rotating neutron stars can be with a high precision approximated by the three parametric Hartle--Thorne solution of Einstein field equations (see  Ref. \refcite{har-tho:1968}, \refcite{ber-etal:2005}). It is known that in most situations modeled with the present neutron star equations of state (EoS) the neutron star external geometry is very different from the Kerr geometry.\cite{Kerr} However, the situation changes when the neutron star mass approaches maximum for a given EoS. For large masses the quadrupole moment does not induce large differences from the Kerr geometry since $\tilde q=QM/J^2$ takes values close to unity. Nevertheless, in general, this does not mean that one can easily neglect the quadrupole moment, since the mass of an average neutron star could be smaller than the maximum mass. For this reason in this work we extend the analyses of Ref. \refcite{ste-vie:1999,ste-etal:1999} involving the Hartle-Thorne solution.

\begin{figure}[t]
\centerline{\includegraphics[width=0.75\columnwidth,clip]{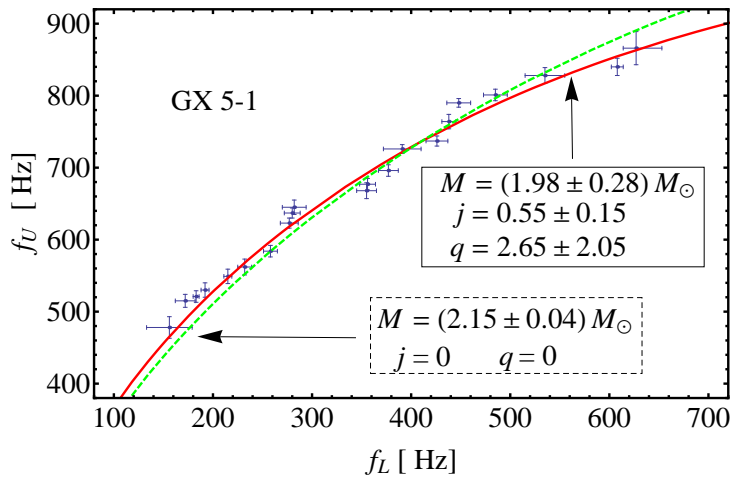}}
\caption{The upper frequency $f_U$ is plotted versus the lower frequency  $f_L$ for the Z source GX 5-1. The dashed green line corresponds to the static case and the solid red line corresponds to the rotating case.}\label{fig:gx51}
\end{figure}
In Fig.~\ref{fig:gx51} we show best fits for the upper frequency versus the lower frequency for the Z source GX 5-1 as it was first demonstrated in Ref. \refcite{boshkayevqpos2014}. We performed fits with all three parameters and with one parameter, the mass $M$. Eventually, the three parameter fit yields $M=(1.98\pm0.28)M_{\odot}$, $j=0.55\pm0.15$ and $q=2.65\pm2.05$. In  terms of physical units the angular momentum and quadrupole moments are $J=(4.66\pm1.83)\times10^{10}$cm$^2$ and $Q=(8.78\pm7.76)\times10^{44}$g$\times$cm$^2$, respectively. Similar analyses have been performed in the Hartle-Thorne spacetime for another source (see Refs. \refcite{stuchlik2014,stuchlik2015} for details). It is worth noticing that the mass of GX 5-1 is unknown from observations. Nonetheless, our inferred mass of a central compact object is consisted with the observed masses of neutron stars in similar X-Ray/Optical binaries (for details, see Ref. \refcite{lattimer2010}).

\begin{figure}[t]
\centerline{\includegraphics[width=0.75\columnwidth,clip]{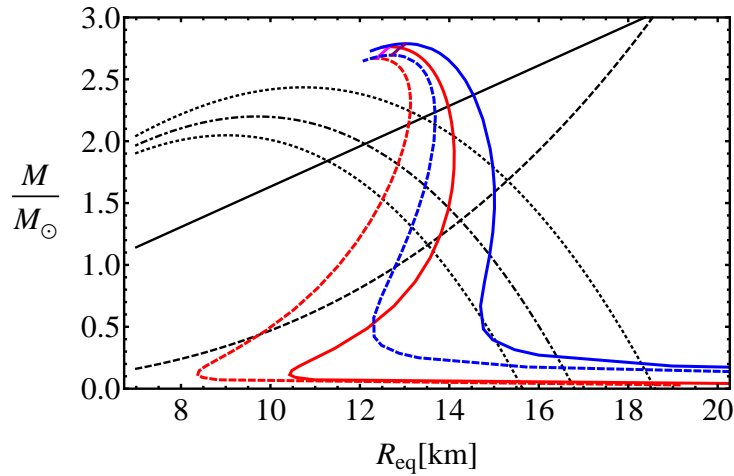}}
\caption{ Observational constraints on the mass-radius relation given by Tr{\"u}mper \cite{trumper2011} and the theoretical mass-radius relation presented in Belvedere et. al\cite{belvedere2014} in Figs. 11 and 24. The red lines represent the configuration with global charge neutrality, while the blue lines represent the configuration with local charge neutrality. The magenta line and the purple line represent the secular axisymmetric stability boundaries for the globally neutral and the locally neutral case, respectively. The red and blue solid lines represent the Keplerian sequences and the red and blue dashed lines represent the static cases. The solid black line is the upper limit of the surface gravity of XTE J1814-338, the  dotted-dashed black curve corresponds to the lower limit to the radius of RX J1856-3754, the dashed black line is the constraint imposed by the fastest spinning pulsar PSR J1748-2246ad, and the dotted curves are the 90$\%$ confidence level contours of constant $R_\infty$ of the neutron star in the low-mass X-ray binary X7. Any mass-radius relation should pass through the area delimited by the solid black, the dashed black and the dotted black lines and, in addition, it must have a maximum mass larger than the mass of PSR J0348+0432, $M=2.01 \pm 0.04 M_\odot$.} \label{fig:MR_constr}
\end{figure}
%
%
\section{Observational and theoretical constraints}\label{sec:4}

It has been pointed out that the most recent and stringent constraints to the mass-radius relation of neutron stars are provided from observational data for pulsars by the values of the largest mass, the largest radius, the highest rotational frequency, and the maximum surface gravity.\cite{trumper2011}

So far, the largest neutron star mass measured with a high precision is the mass of the 39.12 millisecond pulsar PSR J0348+0432, $M=2.01 \pm 0.04 M_\odot$.\cite{antoniadis2013} The largest radius is given by the lower limit to the radius of RX J1856-3754, as seen by an observer at infinity $R_\infty = R [1-2GM/(c^2 R)]^{-1/2} > 16.8$ km,\cite{trumper2004}; it gives the constraint $2G M/c^2 >R-R^3/(R^{\rm min}_\infty)^2$, where $R^{\rm min}_\infty=16.8$ km. The maximum surface gravity is obtained by assuming a neutron star of $M=1.4M_\odot$ to fit the Chandra data of the low-mass X-ray binary X7, it turns out that the radius of the star satisfies $R=14.5^{+1.8}_{-1.6}$ km, at 90$\%$ confidence level, corresponding to $R_\infty = [15.64,18.86]$ km, respectively (see Ref. \refcite{heinke2006} for details). The maximum rotation rate of a neutron star has been found to be $\nu_{\rm max} = 1045 (M/M_\odot)^{1/2}(10\,{\rm km}/R)^{3/2}$ Hz.\cite{lattimer2004} The fastest observed pulsar is PSR J1748-2246ad with a rotation frequency of 716 Hz,\cite{hessels2006} which results in the constraint $M \geq 0.47 (R/10\,{\rm km})^3 M_\odot$. In Fig.~\ref{fig:MR_constr} we show all these constraints and the mass-radius relation presented in Refs. \refcite{belvedere2012} and \refcite{belvedere2014}. 

Similarly to what presented in Belvedere et al. \cite{belvedere2012, belvedere2014} for static and uniformly rotating neutron stars, and introduced by Tr{\"u}mper\cite{trumper2011} the above observational constraints show a preference on stiff EoS that provide largest maximum masses for neutron stars. Taking into account the above constraints, the radius of a canonical neutron star of mass $M = 1.4M_{\odot}$ is strongly constrained to $R\gtrsim12$ km, disfavoring at the same time strange quark matter stars. It is evident from Fig. \ref{fig:MR_constr} that mass-radius relations for both the static and the rotating case presented here, are consistent with all the observational constraints.

\begin{figure}[t]
\centerline{\includegraphics[width=0.75\columnwidth,clip]{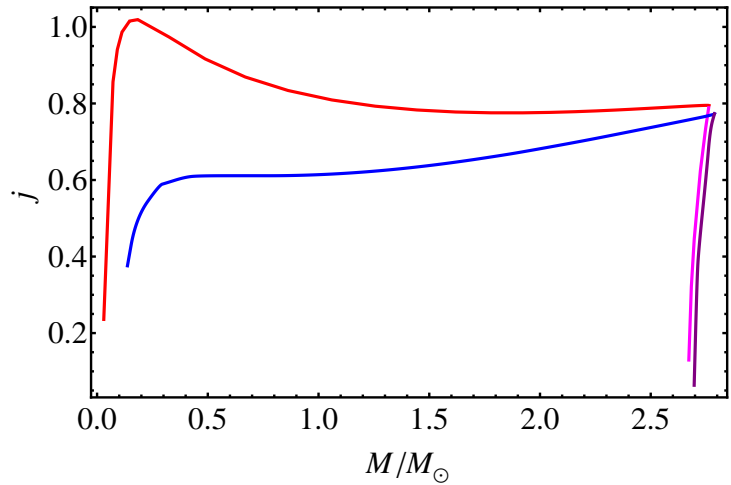}}
\caption{Dimensionless angular momentum versus total mass. Red and blue solid curves are the Keplerian sequences, and magenta and purple curves are axisymmetric secular instability boundaries of both global and local neutrality cases, respectively. }\label{fig:jM_lgcn}
\end{figure}
As for the theoretical constraints, one needs to take into account one more parameter: the dimensionless angular momentum $j$ (spin parameter). Relatively recently Lo \& Lin \cite{lolin2011} revealed that the maximum value of the dimensionless angular momentum $j_{max}$ of a neutron star uniformly rotating at the Keplerian frequency has an upper bound of about 0.7, which is essentially independent on the mass of neutron star as long as the mass is larger than about $1M_{\odot}$. However, the spin parameter of a quark star does not have such a universal upper bound and could be larger than unity.

Qi et. al \cite{qi2014} extended the analyses of Lo \& Lin \cite{lolin2011} considering different kinds of uniformly rotating compact stars, including the traditional neutron stars, hyperonic neutron stars and hybrid stars. It was shown that the crust structure was a key factor to determine the properties of the spin parameter of the compact stars. When the crust EoSs are considered, $j_{max}\sim 0.7$ for $M > 0.5M_{\odot}$ is satisfied for three kinds of compact stars, no matter what the composition of the interior of the compact stars was. When the crust EoSs are not included, the $j_{max}$ of the compact stars can be larger than $0.7$ but less than about $1$ for $M > 0.5M_{\odot}$. Consequently, according to Qi et. al \cite{qi2014} the crust structure provide the physical origin to the stability of $j_{max}$ but not the interior of the compact stars. The strange quark stars with a bare quark-matter surface are the unique one to have $j_{max} > 1$. Thus, one can identify the strange quark stars based on the measured $j > 1.0$, while measured $j\in(0.7, 1.0)$ could not be treated as a strong evidence of the existence of a strange quark star any more.

We show in Fig. \ref{fig:jM_lgcn} the spin parameter versus total mass. Clearly, the value of $j$ is different from those of Lo \& Lin \cite{lolin2011} since we used different approach and different EoS. Despite this the behavior of the $j$ is more similar to those ones of Qi et. al\cite{qi2014} than Lo \& Lin,\cite{lolin2011} as we have crust in both local and global neutrality cases.
%
%
\begin{figure}[t]
\centerline{\includegraphics[width=0.75\columnwidth,clip]{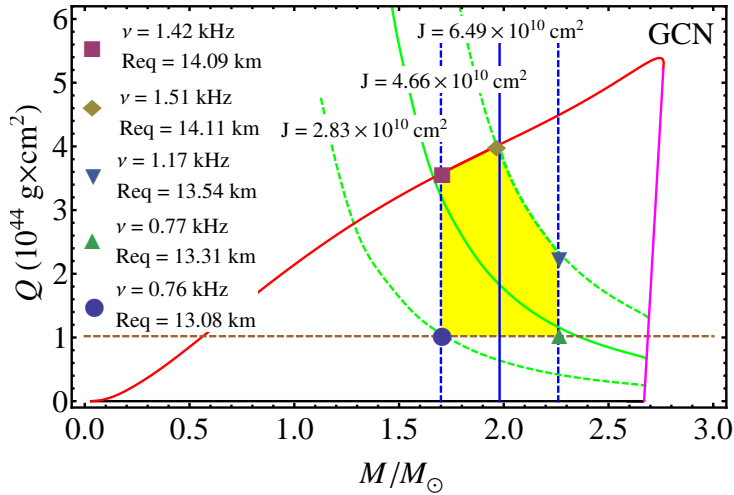}}
\caption{Quadrupole moment versus mass in the global charge neutrality case. The solid red  curve is the Keplerian sequence, the solid purple  curve is the axisymmetric secular instability sequence and the solid black curve on the bottom is the static sequence. A region enclosed in with these curves is called a stability region. All rotating stable configurations can exist only inside the region.  Extracted values of $M$, $j$ and $q$ are plotted in the stability region of rotating neutron stars. Crossing points of these values determine the range of real values. Here we use NL3 model.\cite{belvedere2012}}\label{fig:qmgcn}
\end{figure}
%
%
\section{Estimation of the main parameters of neutron stars}\label{sec:5}
\begin{figure}[t]
\centerline{\includegraphics[width=0.75\columnwidth,clip]{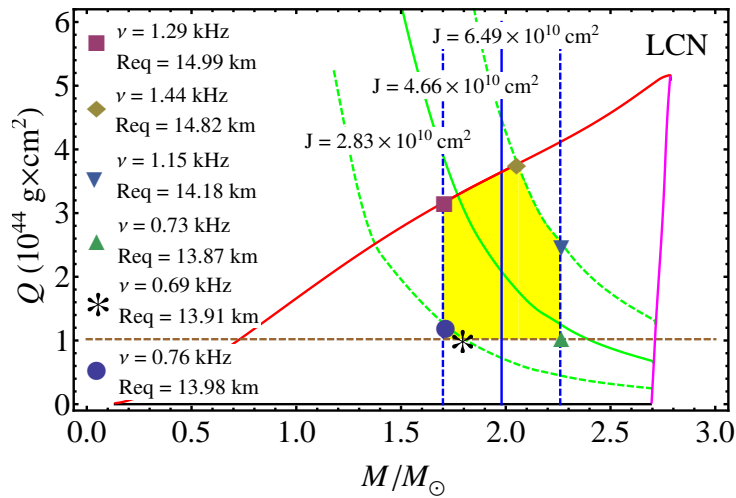}}
\caption{Quadrupole moment versus mass in the local charge neutrality case. Extracted values of $M$, $j$ and $q$ are plotted in the stability region of rotating neutron stars. Crossing points of these values determine the range of real values. Here we use NL3 model.}\label{fig:qmglcn}
\end{figure}

So far, we inferred the $M$, $J$ and $Q$ from the observed QPO data for GX 5-1, interpreting epicyclic frequencies as observed QPOs. We constructed the mass-radius, mass-dimensionless angular momentum and quadrupole moment-mass relations from theory. Now we shall use the inferred $M$, $J$ and $Q$ to estimate the equatorial radius $R_{eq}$ and rotation frequency $f$. Knowing $J$ and $f$ one can easily find the corresponding moment of inertia. In Figs. \ref{fig:qmgcn} and \ref{fig:qmglcn} we show the main parameters of neutron stars for both global and local neutrality cases. 

We plot the Keplerian sequence and secular instability curve in the $Q-M$ diagram. Moreover, we draw curves corresponding to the inferred $M$, $J$ and $Q$ and their error bars, which determine the bounds of $M$, $J$ and $Q$. The crossing points of these bounds with each other and with the stability region indicate the bounds for $R_{eq}$, $f$ and other parameters for GX 5-1 (see the shaded region in Figs. \ref{fig:qmgcn} and \ref{fig:qmglcn}). As we can see the maximum value of the equatorial radius is around 15km and as we expected it is smaller than the innermost radius $r=21$km of the accretion disk and the radius of the marginally stable circular geodesics $r=19.3$km.\cite{boshkayevqpos2015}
%
\section{Conclusion}\label{sec:6}
In this work with the help of the epicyclic frequencies of test particles in the Hartle-Thorne spacetime and the relativistic precession model we have interpreted the quasi-periodic oscillations of the low-mass X-ray binaries. From the observational data of GX 5-1 by fitting fundamental frequencies we extracted the total mass $M$, angular momentum $J$ and quadrupole moment $Q$ of the source with their error bars.

From the fact that $j^2\neq q$ and $j_{max}\leq1$ we came to the conclusion that the source was a neutron rather than a black hole or a strange quark star. On the basis of Ref. \refcite{belvedere2014} we calculated its main parameters by imposing observational and theoretical constraints. As a results by combining theory of neutron stars and observations of quasi-periodic oscillations, we computed the range of basic parameters of GX 5-1 such as the total mass, angular momentum, quadrupole moment, equatorial radius, rotation frequency etc. which play an important role in the physics of neutron stars. For better analyses one needs to consider more sources with refined data. 

\section*{Acknowledgments}

This work was supported by Grant No. F.0679 of the Ministry of Education and Science of the Republic of Kazakhstan. B.K. acknowledges ICRANet for support and hospitality.



\begin{thebibliography}{99}

\bibitem{boshkayevqpos2014} K.~Boshkayev,  D.~Bini,  A.~Geralico, M.~Muccino, J.A.~Rueda,  Gravitation and Cosmology {\bf 20} (4), 233-239 (2014).

\bibitem{ste-vie:1998} L.~Stella,  M.~Vietri, in Abstracts of the 19th Texas Symposium on Relativistic Astrophysics and Cosmology, ed. J. Paul, T. Montmerle \& E. Aubourg (CEA Saclay) (1998).
\bibitem{ste-vie:1999} L.~Stella, M.~Vietri, Phys. Rev. Lett.  {\bf 82}, 17 (1999).
\bibitem{ste-etal:1999} L.~Stella, M.~Vietri, S.M.~Morsink, The Astrophysical Journal, {\bf 524}, 1, L63-L66 (1999).
%
\bibitem{boshkayevqpos2015}	K. Boshkayev, M. Muccino and J. A. Rueda,  Astronomy Reports {\bf 59}, 441 (2015).
%

\bibitem{2010ApJ...714..748T} G.~T{\"o}r{\"o}k, P.~Bakala, E.~{\v S}r{\'a}mkov{\'a}, Z.~Stuchl{\'{\i}}k, M.~Urbanec, Astrophysical Journal {\bf 714}, 748-757 (2010).
%
\bibitem{belvedere2012} R.~Belvedere, D.~Pugliese, J.A.~Rueda, R.~Ruffini, S.-S.~Xue. Nuclear Physics A, {\bf 883}, 1–24, (2012).
\bibitem{belvedere2014} R.~Belvedere, K.~Boshkayev, J.A.~Rueda, R.~Ruffini, Nuclear Physics A {\bf 921}, 33 (2014).
%
\bibitem{boshkayev2012} K.~Boshkayev, R.~Ruffini, H.~Quevedo, Phys. Rev. D {\bf 86} (6), 064043 (2012).


\bibitem{har-tho:1968} J.B.~Hartle, K.S.~Thorne, Astrophysical Journal {\bf 153}, 807 (1968).
\bibitem{Kerr} R.P.~Kerr, Phys. Rev. Letters, {\bf 11}, 237 (1963).
\bibitem{ber-etal:2005} E.~Berti, F.~White, A.~Maniopoulou, M.~Bruni, Monthly Notices of the Royal Astronomical Society, {\bf 358},3, 923-938 (2005).
%
\bibitem{lattimer2010} J. M.~Lattimer, M.~Prakash, (2010). arXiv:1012.3208v1     
%
\bibitem{stuchlik2014} Z. Stuchlik, A. Kotrlova, G. Torok and K. Goluchova,  {\it Acta Astronomica} {\bf 64}, 45 (2014).
%
\bibitem{stuchlik2015} Z. Stuchlik, M. Urbanec, A. Kotrlova, G. Torok and K. Goluchova, {\it Acta Astronomica} {\bf 65}, 169 (2015).
%
\bibitem{trumper2011} J.E.~Tr{\"u}mper, Prog. Part. Nucl. Phys. {\bf 66}, 674–680, (2011).
%

\bibitem{antoniadis2013} J.~Antoniadis et. al, Science {\bf 340} (6131), 448, (2013).

\bibitem{bhattacharyya2005} S.~Bhattacharyya, T.E.~Sthrohmayer, M.C.~Miller, C.B.~Markwardt, The Astrophysical Journal {\bf 619}, 483, (2005).
\bibitem{trumper2004} J.~Tr{\"u}mper, V.~Burwitz, F.~Harberl,  V.E.~Zavlin, Nucl. Phys. B Proc. Suppl. {\bf 132}, 560, (2004).
\bibitem{heinke2006} C.O.~Heinke, G.B.~Rybicki, R.~Narayan, J.E.~Grindlay, Astroph. J. {\bf 644}, 1090, (2006).
\bibitem{lattimer2004} J.M.~Lattimer, M.~Prakash, Science {\bf 304}, 536, (2004).
\bibitem{hessels2006} J.W.T.~Hessels, S.M.~Ransom, I.H.~Stairs, P.C.C.~Freire, V.M.~Kaspi, F.~Camilo, Science {\bf 311},1901, (2006).

\bibitem{lolin2011} K.-W.~Lo, L.-M.~Lin. The Astrophysical Journal, {\bf 728}, 12, (2011).
\bibitem{qi2014} B.~Qi, N.B.~Zhang, B.Y.~Sun, S.Y.~Wang, J.H. Gao, (2014). arXiv:1408.1654v1

%
%

\end{thebibliography}
\end{document}